\documentclass[11pt,titlepage]{article}

\usepackage{times}
\usepackage{amsmath}
\usepackage{amsfonts}
\usepackage{bm}
\usepackage{graphicx}
\usepackage{subfigure}
\usepackage{enumerate}
\usepackage[numbers,round,sort&compress]{natbib}

\newif\ifFigsAtEnd
\FigsAtEndfalse  

\title{Nonlinear Finite Element Analysis of Nanoindentation of Viral Capsids}

\author{Melissa~M.~Gibbons and William~S.~Klug\thanks{ 
           Corresponding author.  Address: 
           Department of Mechanical and Aerospace Engineering,
	   University of California, Los Angeles,
	   420 Westwood Plaza, Room 48-121B,
	   Los Angeles, CA~90095, U.S.A.,
	   Tel.:~(310)794-7347,
  	   Email: klug@ucla.edu }\\
Department of Mechanical and Aerospace Engineering \\ 
University of California Los Angeles,
Los Angeles, CA
}

\date{\today}

\begin{document}

\maketitle

\abstract{
Recent Atomic Force Microscope (AFM) nanoindentation experiments
measuring mechanical response of the protein shells of viruses have
provided a quantitative description of their strength and elasticity.
To better understand and interpret these measurements, and to
elucidate the underlying mechanisms, this paper adopts a
course-grained modeling approach within the framework of
three-dimensional nonlinear continuum elasticity.  Homogeneous,
isotropic, elastic, thick shell models are proposed for two capsids:
the spherical Cowpea Chlorotic Mottle Virus (CCMV), and the
ellipsocylindrical bacteriophage $\phi 29$.  As analyzed by the finite
element method, these models enable parametric characterization of the
effects of AFM tip geometry, capsid dimensions, and capsid
constitutive descriptions.  
The generally nonlinear force response of capsids to indentation is
shown to be insensitive to constitutive details, and greatly
influenced by geometry.  Nonlinear stiffening and softening of the
force response is dependent on the AFM tip dimensions and shell
thickness.  
Fits of the models capture the roughly linear behavior observed in
experimental measurements and result in estimates of Young's moduli of
$\approx$280--360 MPa for CCMV and $\approx$4.5 GPa for $\phi 29$.

\emph{Key words:} viruses; atomic force microscopy; computer simulation; continuum mechanics; elasticity}


\clearpage

\section*{Introduction}

As confirmed by a scan of the RCSB Protein Data Bank \citep{pdb}, the
structural biology techniques of X-ray crystallography and
cryo-electron microscopy (cryo-EM) have enabled determination of the
shape and structure of a multitude of macromolecules, in many cases
locating individual atoms with resolution up to a few angstroms.  In
particular, these methods have been effective in elucidating the
structures of the protein shells (capsids) of many viruses
\citep{BakerEtAl:1999}.
However, despite (or perhaps because of) the wealth of detailed
structural information made available by these methods, it remains a
significant challenge to understand and predict the overall structural
mechanics of capsids based on models of fundamental atomic
interactions.  Global deformations of capsids, such as those
implicated in viral maturation \citep{WikoffEtAl:2006} and infection processes
\citep{ccmv,fhv}, can entail large-scale coordinated motions in which
all atoms are involved.  The shear number of atomic degrees of freedom
in a viral shell makes molecular dynamics simulation of such processes
unreasonable.  In order to understand coarser-scale capsid mechanics,
experimental and theoretical approaches are needed that describe the
capsid structure in reduced terms.

Atomic force microscopy (AFM) is such a tool, able to explicate virus
mechanics by measuring the force response due to indentation of a
capsid.  Recent AFM studies probing the mechanics of viral capsids
\citep{IvanovskaEtAl:2004,Michel} have shown that they can be strong
and yet highly elastic, even under significantly large
deformations.  A particularly interesting hallmark of these
experiments is the \emph{linearity} of force-indentation response.
Prompted by this observation, \citet{IvanovskaEtAl:2004} interpreted
AFM experiments on the $\phi29$ bacteriophage by building theoretical
models based on (linearized) small-strain, thin-shell, continuum
elasticity, thus enabling quantitative estimation of three-dimensional
elastic constants (e.g., the Young's modulus) for the capsid. More
recent experiments have measured a similarly linear force-indentation
response for a completely different virus, Cowpea Chlorotic Mottle
Virus (CCMV) \citep{Michel}, apparently affirming the appropriateness
of linearized elasticity modeling.  However, though linear-response
theory does offer a description of these AFM experiments, it is not at
all clear why a \emph{linear} large-deformation response is observed
in the first place.  Indeed, in traditional solid mechanics,
large deformations are frequently accompanied by a \emph{nonlinear}
force response, even when the material is perfectly elastic.  The
nonlinear elasticity of rubber materials is perhaps the most obvious
example \citep{ogden}.

In this paper the \textit{a priori} assumption of linearity is
abandoned, and more general elastic continuum capsid models are
developed and studied.  Capsids are modeled in the context of
finite-deformation hyperelasticity, wherein strains are not assumed to
be small, such that the nonlinear effects of large displacements,
rotations, and strains are considered.  Previous linearized elasticity
models of capsid indentation \citep{IvanovskaEtAl:2004,lidmar} have
also relied on reduced thin-shell theory \citep{timoshenko,landau},
under which the three-dimensional equations of elasticity are reduced
to two dimensions, facilitating analytical solutions for simple
loadings and geometries.  However, thin-shell theory, as its name
implies, is applicable only for shell structures with a thickness that
is much smaller (by at least a few orders of magnitude) than the
overall structural dimensions.  In particular, thin-shell theory
precludes transverse shear deformation, which though negligible for
thin shells, is an important factor for thick shells.  The nominal
thickness of CCMV is just over 10\% of its outer diameter, likely
putting it outside the range of applicability for thin-shell analysis.
Two options remain for analysis of a thick shell like the CCMV capsid.
The first is to resort to shell theories which allow for shear
deformation \citep{timoshenko}, and the second is to work within the
general framework of 3-D continuum elasticity.  In the present work,
focus is on the latter, more general approach, and its differences
with the former are briefly examined .  Both of these strategies are
difficult to employ in a closed-form analytical manner, and typically
for scenarios involving large deformations and complex geometries or
boundary conditions they require numerical solutions.  The capsid
indentation experiment is one such scenario, involving both large
strains and geometric nonlinearities due to contact between the capsid
and the substrate/AFM.  In this work numerical solution is performed
by the finite element method \citep{Zienkiewicz}, the most popular and
robust analysis technique for elasticity problems in engineering.
Calculations are carried out using ABAQUS \citep{Abaqus}, a
commercially available nonlinear finite element analysis software
package.

Finite element simulations of AFM nanoindentation experiments provide
a means for systematic study of the effects of model parameters on
computed force-indentation response curves, thereby enabling
validation and comparison to experiments.  The inputs to the continuum
elasticity models are of constitutive and geometric natures.  To
understand the influence of constitutive modeling choices, the
classical (Hookean) linear stress-strain law is compared with
nonlinear hyperelastic constitutive laws.  Also, the dependence of
force-indentation curves on capsid dimensions and AFM-tip loading
geometry is demonstrated.  The results show that the nonlinear contact
interaction of the AFM tip with the capsid has important qualitative
and quantitative influence on the structural response.

The experimental studies \citep{IvanovskaEtAl:2004,Michel} that have
motivated this work measured the nanoindentation response of two
viruses: Cowpea Chlorotic Mottle Virus (CCMV) and bacteriophage
$\phi29$.  CCMV is a roughly spherical plant virus of about 28 nm in
diameter.  Its protein capsid, which protects a single-stranded RNA
genome, is assembled from $180$ identical protein subunits arranged
into a truncated icosahedron with a triangulation number $T$=3,
according to the Caspar-Klug classification scheme for icosahedral
viruses \citep{CasparKlug}.
$\phi 29$ is a bacteriophage (i.e., a virus that attacks bacteria)
with a roughly ellipsocylindrical capsid that protects a
double-stranded DNA genome.  The $\phi 29$ capsid is comprised of 235
protein subunits, arranged into a prolate shell with a center
cylindrical region and two icosahedral end caps \citep{TaoOlson1998}.

The following section describes the mechanics framework and the finite
element formulation for the models, and presents the idealized
structural models of CCMV and $\phi29$.  Next, results are presented
of simulations of nanoinentation of CCMV along with some estimates of
effective elastic moduli.  Because its more symmetric shape makes it a
simpler candidate for modeling, CCMV is focused on as the subject of a
series of parametric studies on model parameters.  First the
sensitivity of the model to changes in constitutive laws is
assessed. Despite the presence of large strains in simulations, no
significant difference in overall force-indentation response is
observed for the three different models.  Secondly the influence of
the AFM tip geometry and shell thickness on the overall response is
examined.  These geometric properties are shown to produce both
stiffening and softening nonlinearities in the structural force
response.  For comparison purposes, the computed indentation response
of a $\phi29$ is shown, and an estimate of its effective elastic
moduli is made.

\section*{Models and Methods}\label{sec:model}

\subsection*{Modeling capsid geometry}\label{sec:modelGeometry}

The native atomic structure of CCMV was determined to within $3.2$
\AA\ by \citet{ccmv} using X-ray crystallography and cryo-electron
microscopy.  In this study, the capsid was shown to be a $T=3$
icosahedral shell, composed of protein subunits organized with regions
of five-fold symmetry (pentamers) and six-fold symmetry (hexamers)
that form protrusions from the capsid surface, making the thickness of
the shell highly nonuniform.  For the outer radius of CCMV,
\citet{ccmv} measured a maximum value of
$R^\text{out}_\text{max}=14.2$~nm and a minimum value of
$R^\text{out}_\text{min}=12.0$~nm.  The average outer Radius is
$R^\text{out}_\text{avg}=13.2$~nm \citep{viper}.  For the inner
surface of the capsid they specify a minimum radius of
$R^\text{in}_\text{min}=9.5$~nm and an average radius of
$R^\text{in}_\text{avg}=10.4$~nm.  Here CCMV is idealized as a
uniformly thick spherical shell as shown in
Fig.~\ref{fig:Schematic}(a).  It is not immediately clear what uniform
dimensions for the idealized shell would best represent the highly
nonuniform physical dimensions of the actual capsid.  As a starting
point, average radius values are adopted for the inner and outer
surfaces, which from \citep{ccmv} are $R^\text{in}=10.4$~nm and
$R^\text{out}=13.2$~nm, such that the idealized thickness is $t = 2.8$
nm.  Given the arbitrariness of adopting average dimensions for the
idealized model, a parametric study is done to examine the influence
of these choices on the model results.

The second virus of the study, bacteriophage $\phi 29$, has a less
symmetric but more uniform structure, as determined by
\citet{TaoOlson1998} using cryo-electron microscopy.  Its capsid is
formed by two $T$=3 icosahedral caps connected by a cylindrical region
of hexamers.  The average thickness of the capsid is $1.6$ nm. The two
outer diameters are known to be $54$ nm from the apex of one endcap to
another and $42$ nm across the cylindrical region
\citep{TaoOlson1998}.  The $\phi 29$ capsid is modeled by a center
cylindrical region $24$ nm in length connecting two ellipsoidal
endcaps.  Ellipsoidal endcaps (rather than spherical) are chosen as
there is a disparity between the measured dimension from the center to
the cylindrical section ($21$ nm) and from the center to the apex of
the endcap ($15$ nm).  If the endcaps were made spherical, and the
length of the cylindrical region chosen to match the reported 24 nm
value, then the apex to apex diameter of the model would be much
larger than the reported value.  Model dimensions are summarized in
the schematic diagrams in Fig.~\ref{fig:Schematic}.

\ifFigsAtEnd \relax \else
\begin{figure}[htb]
\centering
\includegraphics[width=4in]{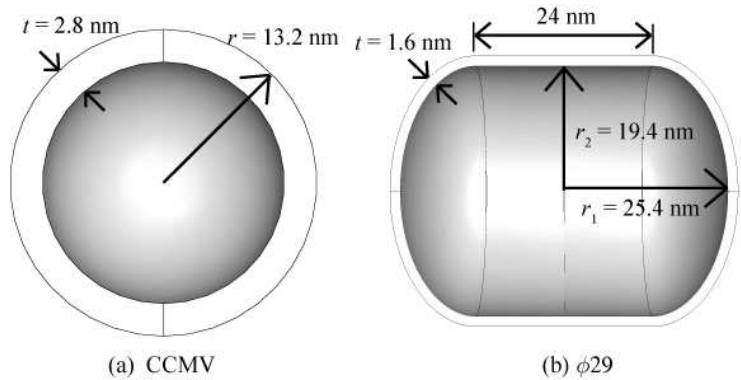}
\caption{\label{fig:Schematic}Schematic of idealized capsid models.
  Dimensions are averaged from the results of structural studies of
  CCMV \protect\citep{ccmv} and $\phi29$
  \protect\citep{TaoOlson1998}.}
\end{figure}
\fi

\subsection*{Theoretical background and computational methodology}
Following the standard definitions of finite deformation continuum
mechanics \citep{Gurtin,Holzapfel}, the deformation of a body is
described by the one-to-one deformation mapping $\bm{\varphi} : {\cal
B}_0 \rightarrow \mathbb{R}^3$, which maps material point at position
$\bm{X}\in{\cal B}_0$ in the reference configuration of the body to
point $\bm{x}=\bm{\varphi}(\bm{X})$ in the deformed configuration of
the body.  The deformation gradient tensor is denoted $\mathbf{F} =
\nabla\bm{\varphi}$, the right Cauchy-Green deformation tensor
$\mathbf{C} = \mathbf{F}^\textsf{T}\mathbf{F}$ and the Green strain
tensor $\mathbf{E} = \frac{1}{2}(\mathbf{C}-\bm{1})$.  The volume
change ratio of current to reference volume is given by the
determinant of the the deformation gradient,
$J=\frac{dv}{dV}=\det\mathbf{F}$.  Also, in the context of
compressible large-strain elasticity, it is helpful to introduce
isochoric (volume-preserving) deformation tensors
$\overline{\mathbf{F}}=J^{-1/3}\mathbf{F}$ and
$\overline{\mathbf{C}}=\overline{\mathbf{F}}^\textsf{T}\overline{\mathbf{F}}$.

Constitutive theory for hyperelastic materials postulates the
existence of a strain energy density function $W$, which
gives the stored elastic energy per unit reference volume at every
material point in the body.  The stress response of a hyperelastic
material is then given in terms of the first Piola-Kirchhoff stress
tensor
$\mathbf{P}=\partial W/\partial\mathbf{F}$ 
or the second Piola-Kirchhoff stress tensor 
$\mathbf{S} = \partial W/\partial\mathbf{E}$.

Mechanical equilibrium can be enforced by minimizaton of the total
mechanical energy (including work of contact forces).  The finite
element method approximated the nonlinear elasticity problem by
interpolating the deformation mapping among the deformed positions of
the vertices (nodes) of a polyhedral mesh, and solving for the nodal
positions which best minimize energy.
The finite element equilibrium equations are nonlinear in the
unknown deformed nodal positions $\bm{x}_a$ because of the
nonlinearities in the constitutive expressions for stress and
(for the present context) in the contact conditions which define
external forces.

\subsection*{Constitutive laws}\label{sec:ConstLaws}
In this work, three particular hyperelastic models are employed: St.\
Venant-Kirchhoff, Neo-Hookean, and Mooney-Rivlin. The so-called
\emph{St.\ Venant-Kirchhoff} model, is given by the strain energy
function
\begin{equation}\label{eq:SVK-W}
W = \frac{\lambda_0}{2} (\text{tr}\,\mathbf{E})^2 
+ \mu_0 (\text{tr}\,\mathbf{E}^2) .
\end{equation}
The chief appeal of this material model is that the resulting (second
Piola) stress response is linear in the Green strain,
\(
\mathbf{S} = \lambda_0(\text{tr}\,\mathbf{E})\bm{1} + 2\mu_0 \mathbf{E} ,
\)
and is, in a sense, akin to Hooke's law for isotropic
(infinitesimal-strain) linear elasticity.  Linearization of this
stress response for vanishingly small values of the displacement
gradient $\mathbf{H}=\mathbf{F}-\bm{1}$ identifies $\lambda_0$ and
$\mu_0$ as the initial Lam\'e constants.

For isotropic compressible hyperelastic materials it can be convenient
to define the strain energy function in a decoupled representation as
\[ 
W = W_\text{vol}(J) + W_\text{iso}(\bar{I}_1, \bar{I}_2) , 
\]
where $ W_\text{vol}(J)$ describes the volumetric response, and
$W_\text{iso}(\bar{I}_1.\bar{I}_2)$ describes the isochoric response
in terms of the first two principal invariants of
$\overline{\mathbf{C}}$.
The simplest example of such a constitutive law, the
\emph{Neo-Hookean} model defines the isochoric response to be simply
linear in the first invariant:
\begin{equation}\label{eq:neoHookean}
W_\text{iso} = \frac{\mu_0}{2}(\bar{I}_1 - 3) .
\end{equation}
Emerging from the statistical mechanics of a Gaussian chain
\citep{Treloar}, the Neo-Hookean form has some justification as a
highly idealized constitutive model for polymers and proteins.
Linearization of the Neo-Hookean stress response yields $\mu_0$ as the
initial shear modulus.
The \emph{Mooney-Rivlin} model can be though of as an extension of the
Neo-Hookean to include the lowest-order dependence on the second
invariant:
\begin{equation}\label{eq:MooneyRivlin}
W_\text{iso} = C_{10}(\bar{I}_1 - 3) + C_{01}(\bar{I}_2 - 3) .
\end{equation}
Linearization of Mooney-Rivlin for small strains gives an initial
shear modulus of $\mu_0=2(C_{10}+C_{01})$.
For both Neo-Hookean and Mooney-Rivlin models, we choose a simple
quadratic volumetric response
\(
W_\text{vol} = k_0(J-1)^2/2 ,
\)
such that linearization renders $k_0=\lambda_0 + \frac{2}{3}\mu_0$ as 
the initial bulk modulus, consistent with initial Lam\'e constants, 
$\lambda_0$ and $\mu_0$.

\subsection*{Simulation of capsid nanoindentation}

AFM nanoindentation of capsids is simulated by the finite element
method, using the commercially available package ABAQUS
\citep{Abaqus}.  The models described above are meshed with finite
elements, and the capsid meshes are compressed between substrate and
AFM tip.  The substrate and AFM tip are both modeled as rigid.  The
substrate has the geometry of a flat plate, whereas the AFM tip is
given a spherical shape with a nominal radius of about 14 nm,
consistent with geometry of the tips used in the experiments of
\citet{IvanovskaEtAl:2004} and \citet{Michel}.  Though the spherical
CCMV model obviously has no intrinsic orientation, the same is not
true for the ellipsocylindrical $\phi29$.  Again motivated by the
experiments of \citet{IvanovskaEtAl:2004}, the $\phi29$ model is
oriented such that compression is simulated along the short axis of
the capsid.  Noting presence of symmetries in the capsid and loading
geometries, it seems reasonable to expect deformation to obey the same
symmetries.  Hence only one-quarter of each capsid is meshed, and
symmetry boundary conditions along the boundaries are enforced.
Meshed model assemblies are shown in Fig.\ \ref{fig:undeformedMesh}.

\ifFigsAtEnd \relax \else
\begin{figure}
  \centering
  \includegraphics[width=4in]{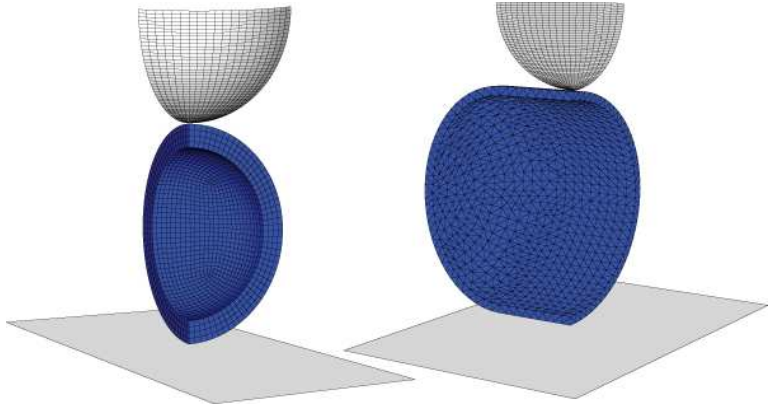}
  \caption{\label{fig:undeformedMesh}Finite element meshes of CCMV
    (left) and $\phi 29$ (right) Capsids.  Dimensions of the capsids
    are shown in Fig.~\protect\ref{fig:Schematic}.  AFM tip is modeled
    as a rigid hemispherical shell with $R=14$ nm, substrate as a
    rigid flat plate. (Capsids are shown not on the same scale.)}
\end{figure}
\fi

Contact of the rigid tip and substrate with the deformable capsid is
modeled as frictionless, such that Lagrange-mulitiplier contact forces
are introduced normal to the contact interfaces, in order to prevent
interpenetration.  The progressive compression of the capsid is
achieved by an incremental, quasi-static, displacement-controlled
process, where each increment involves a small upward displacement of
the rigid substrate followed by iterative solution of the nonlinear finite
element equilibrium equations consistent with contact constraints.
For each incremental displacement of the substrate, the total contact
force at each contact interface is computed, allowing construction of
a force-deflection curve for the indentation process.

\section*{Nanoindentation of CCMV}\label{sec:CCMVresults}
As described in the previous section, parameters defining the
spherical model of CCMV are of two types: geometric and constitutive.
The simulations described here are designed to separately and
quantitatively understand the effects of these parameters on capsid
indentation.  
The influence of capsid dimensions is considered by fixing the radius
of the midsurface of the shell (equidistant from the inner and outer
surfaces) at $R=11.8$nm, while varying the thickness from 1 nm to 5 nm
(roughly reflecting the range of physical thickness over the actual
capsid shell).  A range of AFM tip geometries is also considered, from
a rather sharp tip of radius 3.5 nm to a completely flat tip (i.e.,
infinite radius).

The three consititutive theories described above---St.\
Venant-Kirchhoff, neo-Hookean, and Mooney-Rivlin---are parameterized
in a manner such that they linearize consistently with initial 
Lam\'e constants related to initial Poisson ratio $\nu$
and initial Young's modulus $E$ by
\[
\lambda_0 = \frac{\nu E}{(1-2\nu)(1+\nu)} \qquad \mu_0 = \frac{E}{2(1+\nu)} .
\]
 Informal comparisons showed that the model results were quite
insensitive to the choice of Poisson ratio. Because proteins tend
toward behaving incompressibly in their elasticity, an only slightly
compressible Poisson ratio of $\nu=0.4$ is chosen here for all
simulations.  As can be seen from the definitions earlier, because the
simulation is displacement-controlled, the Young's modulus acts
effectively as a simple proportionality factor for the stress response of the
material (and hence also for the force response of the entire shell).
Therefore, simulation results can be normalized by $E$, avoiding the
need for running a series of simulations over a range of values.

Parametric studies of geometric and constitutive effects are described
in the following section.  As a representative example, consider a
simulation of CCMV with the nominal dimensions from
Fig.~\ref{fig:Schematic} (outer radius of 13.2 nm and a thickness of
2.8 nm), described by St.\ Venant-Kirchhoff elasticity, indented by a
spherical rigid AFM tip of radius 14 nm.  For a capsid with Young's
modulus of $E=250$MPa (on the order of what's expected for typical
protein material \citep{Howard:2001}), Fig.~\ref{fig:StressContourCCMV} shows a sequence of deformed capsid
shapes, color coded by von Mises stress \citep{Ugural} at several
intervals during the indentation process.

\ifFigsAtEnd \relax \else
\begin{figure}
  \centering
  \includegraphics[width=4in]{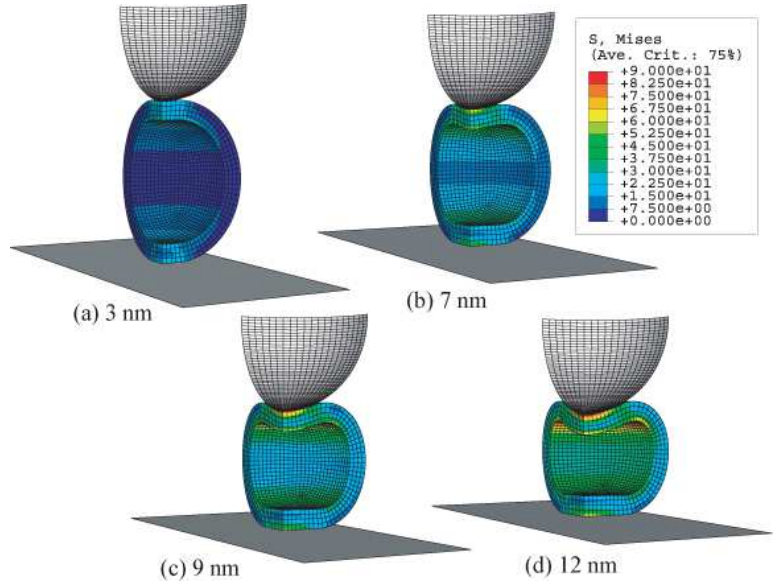}
  \caption{\label{fig:StressContourCCMV}Indentation of spherical model
  of CCMV with dimensions from Fig.~\protect\ref{fig:Schematic}, and a
  constitutive response defined by the St.\ Venant-Kirchhoff theory
  with $E=250$ MPa, and $\nu=0.4$.  Deformed capsid shapes are shown for 
  several values of substrate displacement.  Color contours indicate 
  the von Mises stress.}
\end{figure}
\fi

Fig.~\ref{fig:StressContourCCMV} shows that as AFM indentation
increases during simulation, elastic stresses build and are
distributed through the thickness of the capsid wall (as indicated by
the contours of von Mises stress).  At the outset, the stress is
highest at and around the point of contact between the AFM tip and the
top of the capsid, and is more uniformly distributed through the
thickness away from the contact region.  As the substrate displacement
continues, the stress is increasingly concentrated on the inside surface
of the capsid.  In general at later stages in the indentation, the von
Mises stress decreases in magnitude through the thickness, taking its
lowest value at the center of the capsid wall, consistent with the
interpretation that bending becomes a more dominant load-carrying
mechanism.  For this and all other CCMV simulations, indentation is
carried forward until the substrate displacement reaches 30\% of the
capsid diameter, approximately the point of failure in experiments
\citep{Michel}.  At this point in this representative simulation, the
largest von Mises stress is about 90 MPa and the maximum magnitude of
the principal strains is about 15\%, clearly outside the range of
applicability of small strain linear elasticity.  Since many materials
have ultimate strengths that are in the range of 1--10\% of their
Young's moduli \citep{Howard:2001}, the maximum stress of 90 Mpa far
exceeds this rule of thumb for indicating failure.  This suggests the
capsid is qualitatively similar to rubber materials, which typically
have higher ultimate strengths and lower Young's moduli.  

Given the magnitude of strains, along with the significant changes in
contact geometry throughout the simulation depicted in
Fig.~\ref{fig:StressContourCCMV}, it is somewhat surprising that the
corresponding force-indentation curve, shown in
Fig.~\ref{fig:CCMVforceResponse}, exhibits only very subtle
nonlinearity.  For very small displacements of the substrate, there is
a noticeable positive curvature in the force response.  This
corresponds to the initial part of the indentation, where deformation
gradually spreads through the thickness of the shell as the contact
area between tip and capsid grows.  In the limit of small
deformations, this should lead to a Hertz-like contact force scaling,
which from linear elasticity is expected to take the form $F\sim
d^{3/2}$ \citep{timoshenko,landau}, where $F$ is the total contact
force and $d$ is the indentation or change in capsid height.  This
expected scaling is consistent with initial convexity of the simulated
force response.

\ifFigsAtEnd \relax \else
\begin{figure}
\centering
\includegraphics[width=4in]{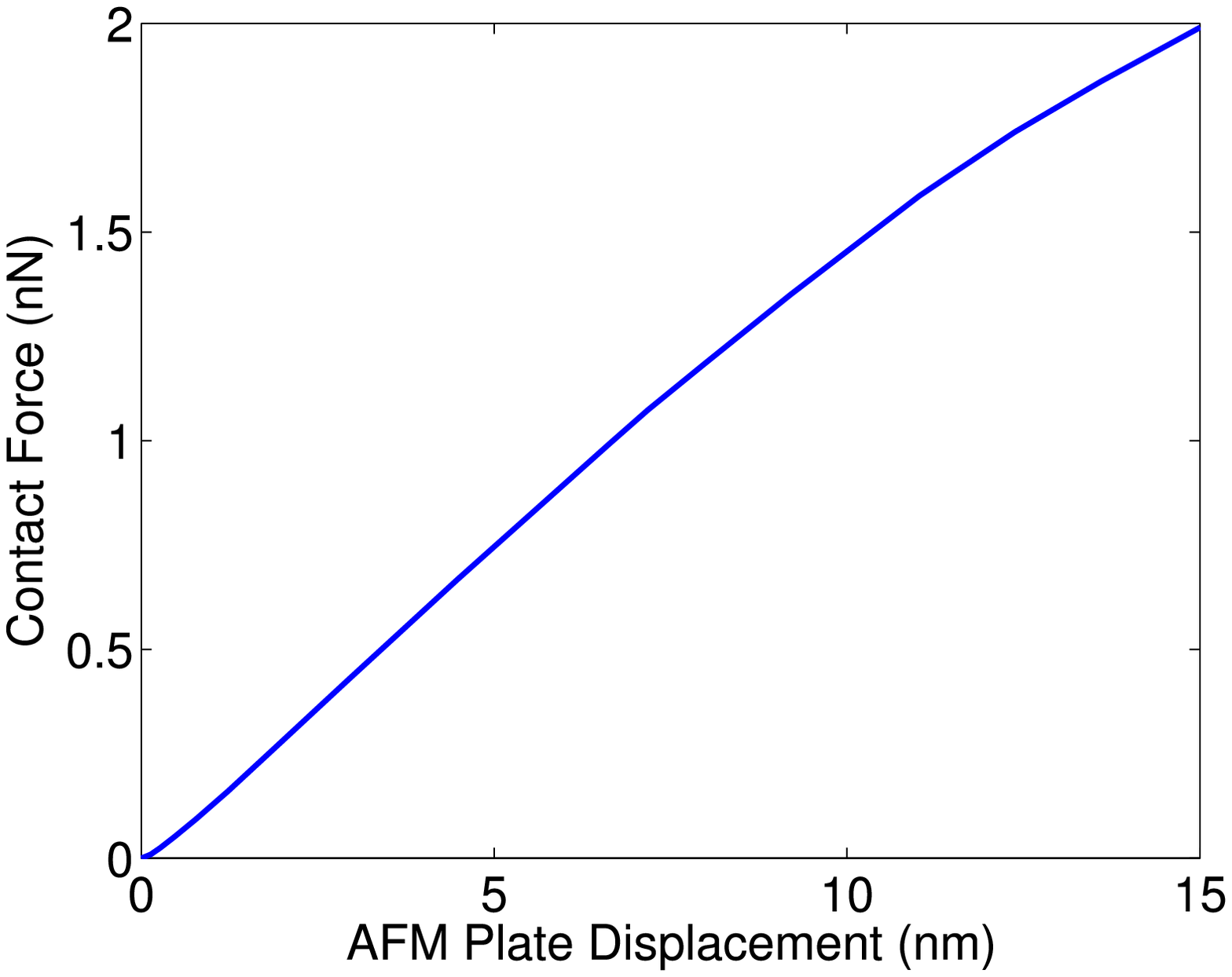}
\caption{\label{fig:CCMVforceResponse} Force response for indentation
of spherical CCMV model from Fig.~\protect\ref{fig:StressContourCCMV}.}
\end{figure}
\fi

At larger indentations, the curve becomes remarkably linear over a
large range, before eventually softening around $d\approx 7-8$ nm as
bending deformations of the shell become more severe.  It is in this
range in the simulation that the contact pressure near the center of
the AFM tip contact region begins to decrease to the point where
eventually full contact is lost and a gap opens between the AFM tip
and the apex of the capsid.  The contact region is from then on
defined by a circular ring.  Fig.~\ref{fig:TipSeparation}(a) shows a
close-up side view of the AFM tip and the top of the capsid
illustrating the separation of the capsid away from the AFM tip.
Fig.~\ref{fig:TipSeparation}(b) shows the contact pressure between the
AFM tip and capsid surfaces at indentations of $d\approx 8$ nm and
$d\approx 9$ nm, just after the capsid and AFM tip separate.  The blue
area in the center of the top portion of the capsid represents zero
contact pressure, and the circular ring area where the AFM tip and
capsid are still in contact can clearly be seen.  Zero contact
pressure is indicative of a loss of contact several nanometers before
the separation (seen in Fig.~\ref{fig:TipSeparation}a) is visible.
The inner gray region in Fig.~\ref{fig:TipSeparation}(a) is the area
of the two surfaces that are no longer in contact.
This separation of the capsid apex from the AFM tip geometrically
resembles the so-called ``snap-through'' buckling of arches and thin
shells \citep{timoshenko,Calladine}.  This nonlinear contact effect is
not to be confused with a geometric instability or equilibrium
bifurcation, which would be marked by a drop in the force-indentation
curve.  However, the softening of the force-indentation curve is
qualitatively similar to ``snap-through'' behavior, and consistent
with a transition of the primary load-bearing mechanism from
stretching to bending of the shell surface.  The absence of a drop in
force for this reversibly elastic model perhaps suggests that the
failure experimentally observed by \citet{Michel} may be related to
instability in the constitutive response of the protein material.

\ifFigsAtEnd \relax \else
\begin{figure}
  \centering
  \includegraphics[width=4in]{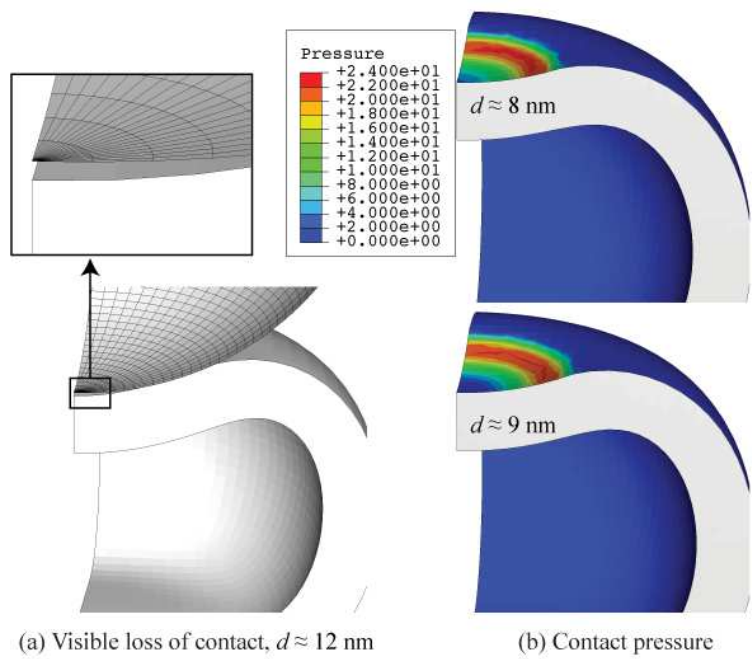}
  \caption{\label{fig:TipSeparation}At higher deformations, the apex
    of the CCMV capsid is observed to separate from the indenting AFM
    tip.  Separation is preceded by a drop in contact pressure near
    the apex.}
\end{figure}
\fi

\subsection*{Parametric studies}

\paragraph{Constitutive laws.}
As shown in the previous section, AFM indentation is capable of
inducing significant strains in a capsid shell.  The
experimentally-demonstrated fact that CCMV capsids remain elastic
despite such deformations, puts them in a very special class of
materials.  Phenomenological constitutive theory for large-strain
nonlinear-elastic materials has traditionally focused on the response
of rubber-like materials.  Two of the lower-order theories for
rubber-like materials: neo-Hookean and Mooney-Rivlin, are employed
here to get a sense for how important constitutive details are
relative to geometric nonlinearities in determining the overall shape
of the force-indentation curve.  Fig.~\ref{fig:NeoHookean} compares
force-indentation responses for capsids modeled with neo-Hookean and
Mooney-Rivlin to that of the linear St.\ Venant-Kirchhoff model.  The
parameters for all of these simulations are identical to those the
representative CCMV model above, with the dimensions of the shell
taken from Fig.~\ref{fig:Schematic}, a Poisson's ratio of $\nu=0.4$
and an initial Young's modulus of $E=250$MPa.  The coefficients of
three-parameter Mooney-Rivlin model are determined by specifying the
fraction $\frac{C_{10}}{C_{01}}$ in addition to $E$ and $\nu$.  Three
values of this fraction are considered, such that the first invariant
is weighted respectively less, evenly, and more than the second
invariant.  Fig.~\ref{fig:NeoHookean} clearly demonstrates that the
force-indentation curves for all of these constitutive models are
practically coincident throughout the entire deformation.  This
clarifies that the details of the constitutive response are not
centrally important to the shape of the force-indentation curve.  It
suggests rather that the geometry of the shell, and the geometry of
loading are more influential in determining the indentation response.
Furthermore, these results support the appropriateness of the linear
stress-strain law.  Apparently, despite the presence of rather large
strains, constitutive nonlinearities are masked from the overall
structural response.  This is likely due to the fact that the strains
over much of the capsid are significantly smaller than the maximum
value, such that locally even the nonlinear models are in the linear
response regime.

\ifFigsAtEnd \relax \else
\begin{figure}
  \centering
  \includegraphics[width=4in]{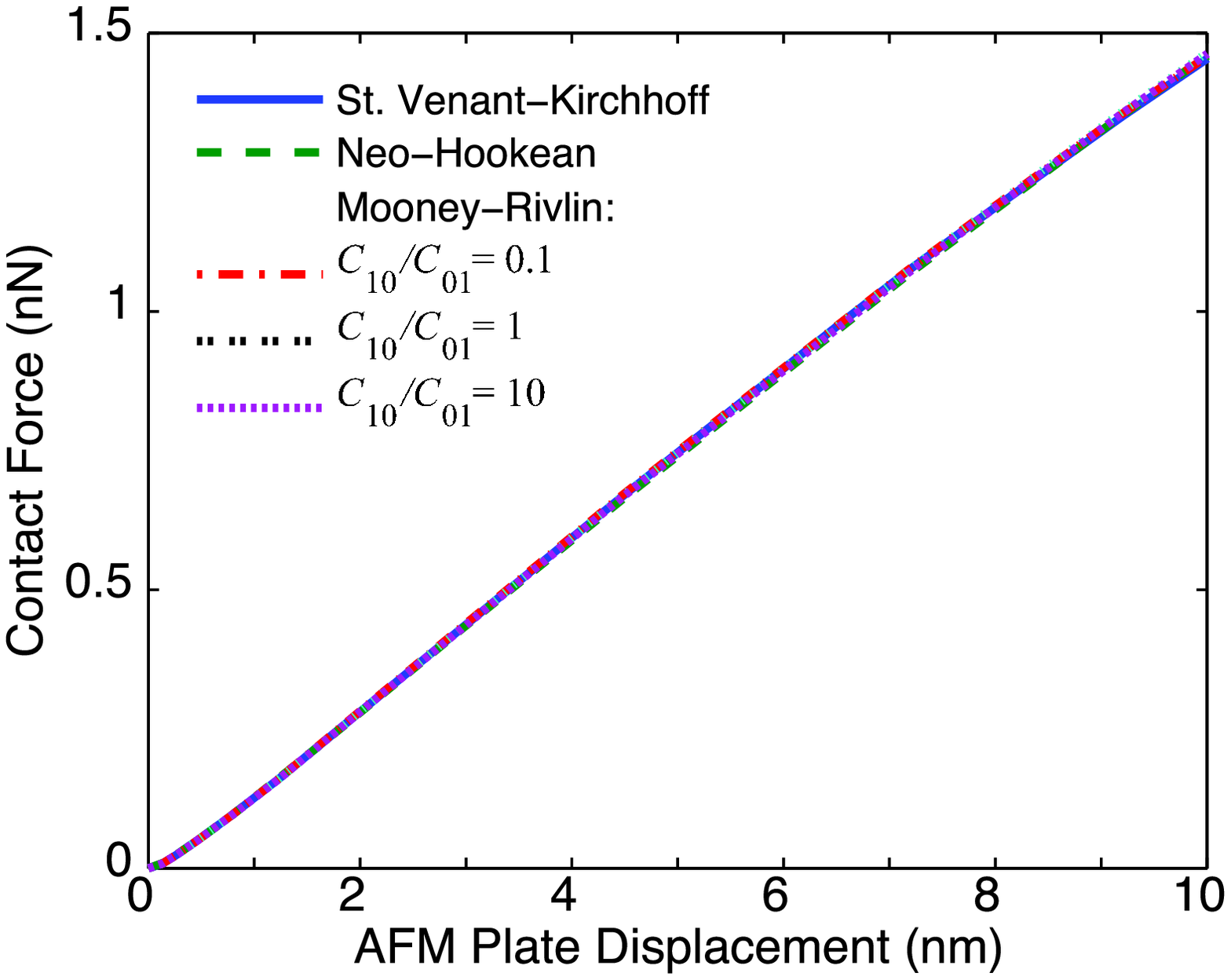}
  \caption{\label{fig:NeoHookean} Sensitivity to constitutive law.
  Contact force curves for St.\ Venant-Kirchhoff, compressible
  Neo-Hookean, and compressible Mooney-Rivlin models.  Representative
  capsid dimensions from Fig.~\protect\ref{fig:Schematic} are used.}
\end{figure}
\fi

\paragraph{AFM tip size.}
In the experiments of \citep{IvanovskaEtAl:2004,Michel} capsids rest
on a flat substrate and are compressed by an AFM tip which is roughly
hemispherical.  The AFM tip radius is on the order of the outer radius of the 
CCMV capsid, but exact determination of the dimensions is
difficult.  Hence there is a need to determine the importance of the
modeled AFM tip size on the resulting contact behavior.  Simulations
were performed varying the AFM tip radius from one-quarter of its
nominal size (14 nm) to infinite (treating the AFM tip as a second
flat plate), and the resulting contact force curves are compared in
Fig.~\ref{fig:AFMTip}.  At displacements of $7-8$ nm, when failure
is experimentally observed, the contact force varies by $\approx 15\%$.
When approaching either extreme in the size of the AFM tip, be it of
infinite radius (a flat plate) or nearing a point load (one quarter of
the initial radius), the results change only slightly.  Also notable
is the subtle change in shape of the force-indentation curve as the
tip radius is changed.  Softening of the force response, noted earlier
to occur along with a separation of the capsid apex from the tip, is
accentuated by smaller tips which induce more bending deformation in
the capsid.  Indeed, over the range of deformation simulated, this
softening is completely absent when the tip is modeled as a flat
plate.  In this case bending deformations near the apices is limited
and the slope of the force-indentation curve is monotonically
increasing.

\ifFigsAtEnd \relax \else
\begin{figure}
  \centering
  \includegraphics[width=4in]{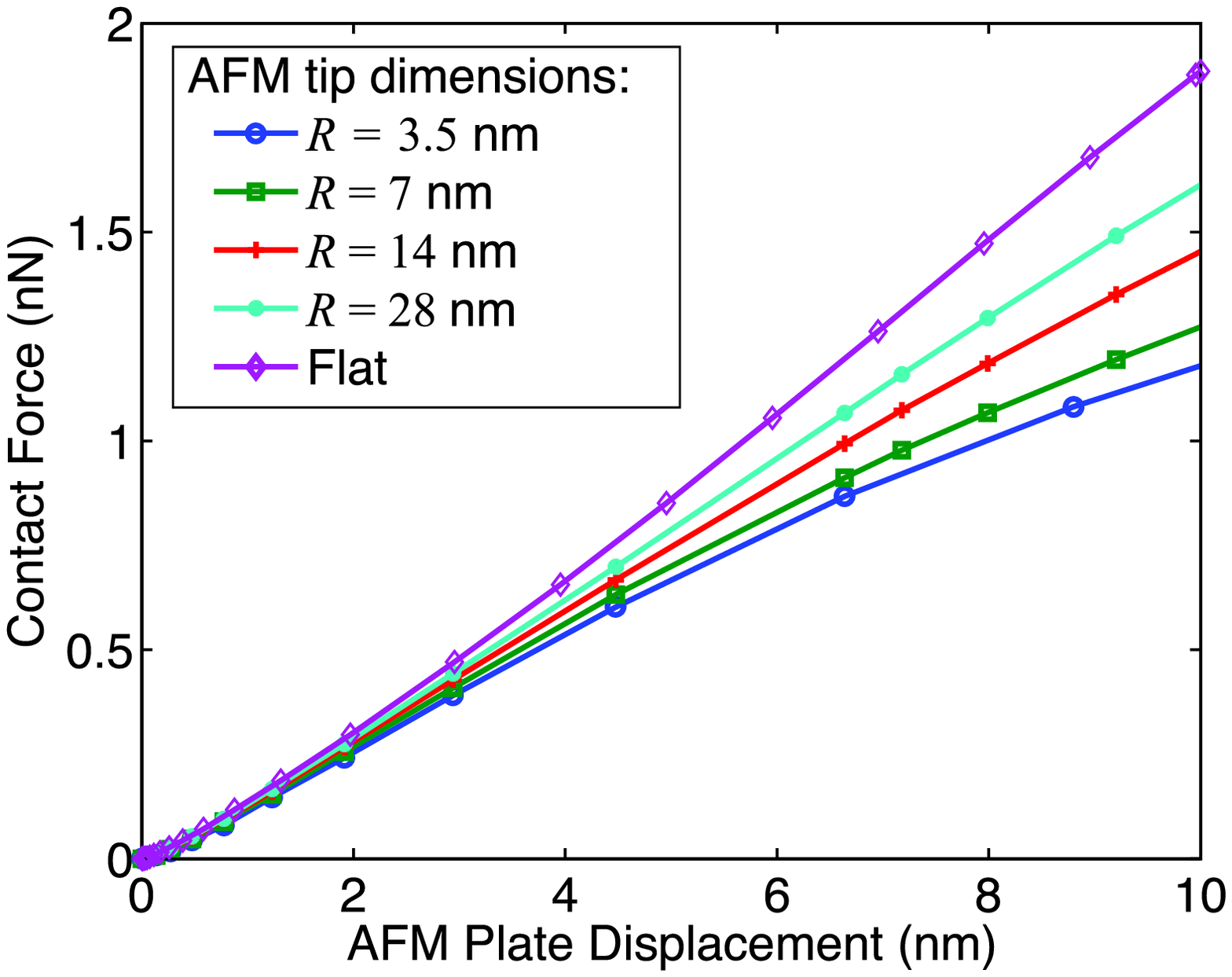}
  \caption{\label{fig:AFMTip}Contact force comparison for several AFM
  tip sizes.  Representative capsid dimensions from
  Fig.~\protect\ref{fig:Schematic} are used.}
\end{figure}
\fi

\paragraph{Capsid thickness.}
As noted earlier, the Young's modulus, $E$, acts as a simple
proportionality factor for stress and force responses to indentation,
allowing for simple renormalization of the force-indentation response
to eliminate dependence on $E$.  The same cannot be said for the
scaling with capsid thickness $t$.  Because nonlinear geometric
effects are included in the finite element model, it is difficult to
predict analytically the precise scaling with thickness.  Roughly,
thickness scaling should follow that of linearized thin-shell theory
wherein force goes as the square of thickness $F\sim t^2$
\citep{timoshenko, landau}, but this is only an approximation.
Fig.~\ref{fig:ForceComparison} clearly shows the change in behavior as
the thickness of the capsid is varied, while holding the average
radius constant at $11.8$ nm.

\ifFigsAtEnd \relax \else
\begin{figure}
  \centering
  \includegraphics[width=4in]{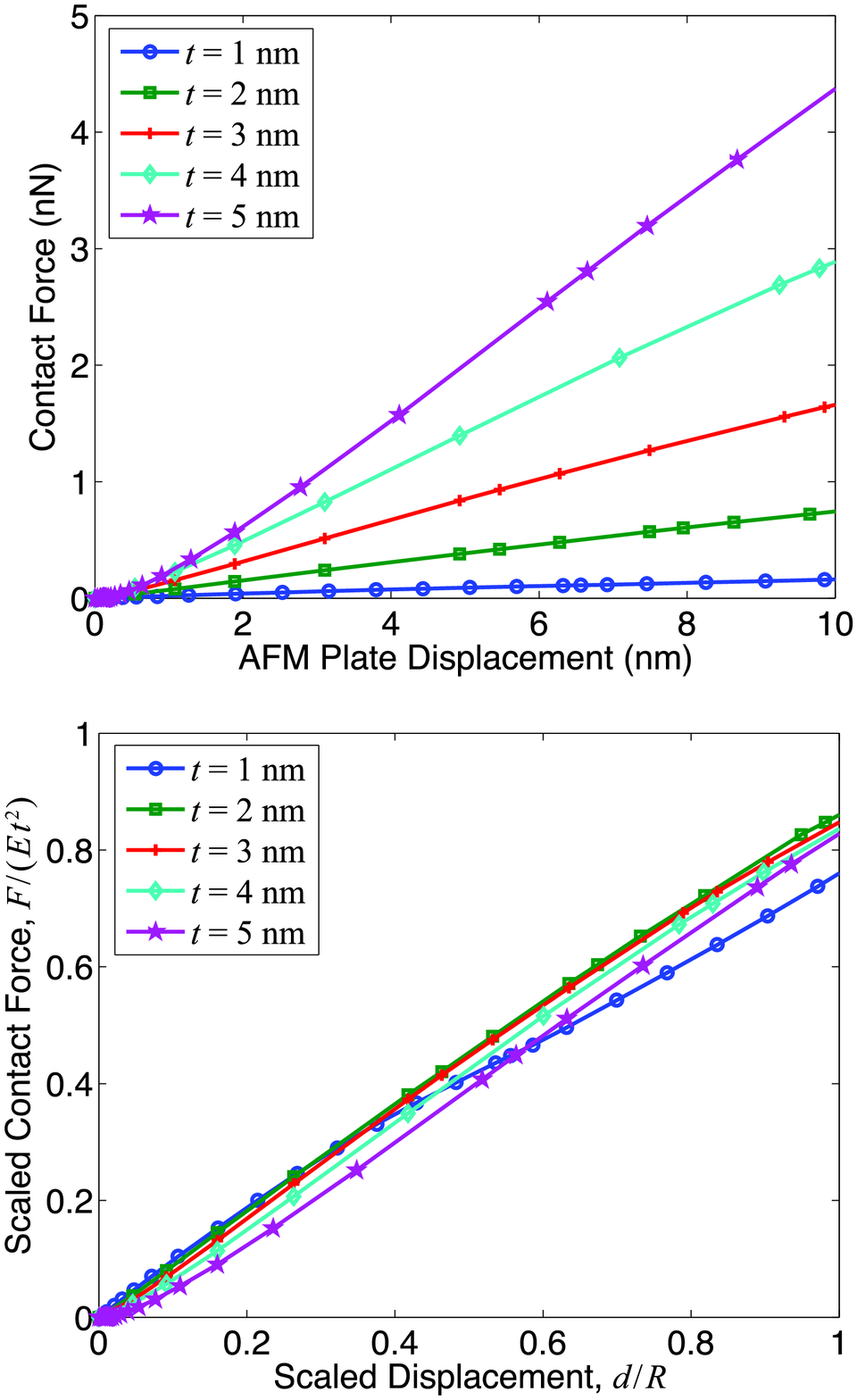}
  \caption{\label{fig:ForceComparison} Contact force response of CCMV
  over a range of capsid thicknesses.  (Top) Force vs. AFM plate
  displacement for $E=250$ MPa and $R_{avg}=11.8$
  nm. (Bottom) Dimensionless renormalization of force response
  according to expected scaling from linearized thin-shell
  elasticity.}
\end{figure}
\fi
 
For deformation of a thin shell, \emph{linearized} elasticity theory
gives a force displacement scaling
$F\propto {Et^2}d/R$,
where $t$ is the shell thickness, $R$ is the size of the shell, and
$E$ is the Young's modulus \citep{timoshenko, landau}.  In general the
proportionality factor is a function of the loading and boundary
conditions of the problem.  For the present application,
Fig.~\ref{fig:ForceComparison}(b) plots the the contact force,
normalized by $Et^2$, versus indentation, normalized by CCMV's average
radius $R$.  Even though this model considers geometric
nonlinearities, the curves for different thicknesses nearly collapse
to the linearized scaling, with an average proportionality factor of
0.8, i.e.,
\[
F\approx 0.8\frac{Et^2}{R}d .
\]
Assuming a nominal thickness of $t=2.8$nm and average radius of
$R=11.8$nm for CCMV, the Young's modulus can be estimated by relating
the experimentally measured spring constant of the capsid to 
the approximate theoretical spring constant of $0.8Et^2/R$.  In this
way, the Young's modulus is calculated to be $280$ MPa for the empty
wild-type capsids and $360$ MPa for the empty subE mutant capsids.
Notably, these modulus values are similar to those of soft plastics,
such as Teflon, and are slightly smaller than characteristic values for
single proteins as measured in single molecule experiments
\citep{Howard:2001}.  This is consistent with the notion that much of
the local deformation of actual capsids may be sustained in the
regions between protein subunits, such that the overall stiffness of
the capsid structure is less than that of the individual proteins.

The renormalized plot in Fig.~\ref{fig:ForceComparison}(b) also
reveals more clearly how deviations from linearity in the force
response depend on shell thickness.  For larger thicknesses, the
stiffening Hertz-like response near the origin is more pronounced,
while for smaller thicknesses softening of the force response at
larger indentations is more prominent.  The mechanisms for both of
these nonlinear effects are related to changes in the geometry of
contact between the AFM tip and the capsid shell, which are primarily
controlled by the ease with which the shell can be bent to
accommodate the rounded tip.  Notably, for thicknesses near the
average physical thickness of the CCMV capsid (2.8 nm), these two
nonlinear effects (one stiffening and the other softening) seem to
balance each other out resulting in a nearly linear response.  It is
perhaps reasonable to conjecture that these mechanisms have roles in
producing the linearity observed experimentally for the actual capsid.

\section*{Nanoindentation of $\phi 29$}\label{sec:phi29}

A model assembly similar to that applied to CCMV above has been
employed to simulate indentation of $\phi 29$, mimicking the
experiments of \citet{IvanovskaEtAl:2004}.  As the same experimental
setup was involved in experiments on both capsids, here the AFM tip
geometry is again modeled as spherical with a radius of 14 nm.  However,
$\phi29$ has a significantly larger capsid than CCMV, by a factor of
two in diameter.  Hence, the same 14-nm AFM tip is smaller relative to
$\phi29$.  In addition, the nominal thickness of the $\phi29$ capsid
relative to its average radius is much smaller than that of CCMV (by a
factor of four).  Recalling that softening of the force response for
CCMV was accentuated both by smaller tip size and smaller shell
thickness, it is not surprising to see in
Fig.~\ref{fig:ForceComparisonPhi29} that the simulated
force-indentation response of $\phi29$ exhibits more significant
softening at moderate to large indentations over a range of modeled
thicknesses.  The response of $\phi29$ adheres reasonably well to the
linearized scaling, however the shell-thickness-related softening
nonlinearity appears more pronounced than for CCMV.

From Fig.~\ref{fig:Schematic} the nominal capsid thickness $t=1.6$ nm,
and a nominal radius (chosen to be the average value between the
smaller and larger radii, $R=23.2$ nm) are chosen such that an estimate of the linearized
proportionality factor, and capsid Young's modulus can be made.  From
Fig.~\ref{fig:ForceComparisonPhi29}, for a wide range of
thicknesses, a good estimate of the average proportionality factor is
$0.6$.  Thus, from the experimentally determined spring constant of
the $\phi29$ capsid, the Young's modulus is estimated to be $E=4.5$
GPa.

\ifFigsAtEnd \relax \else
\begin{figure}
  \centering
  \includegraphics[width=4in]{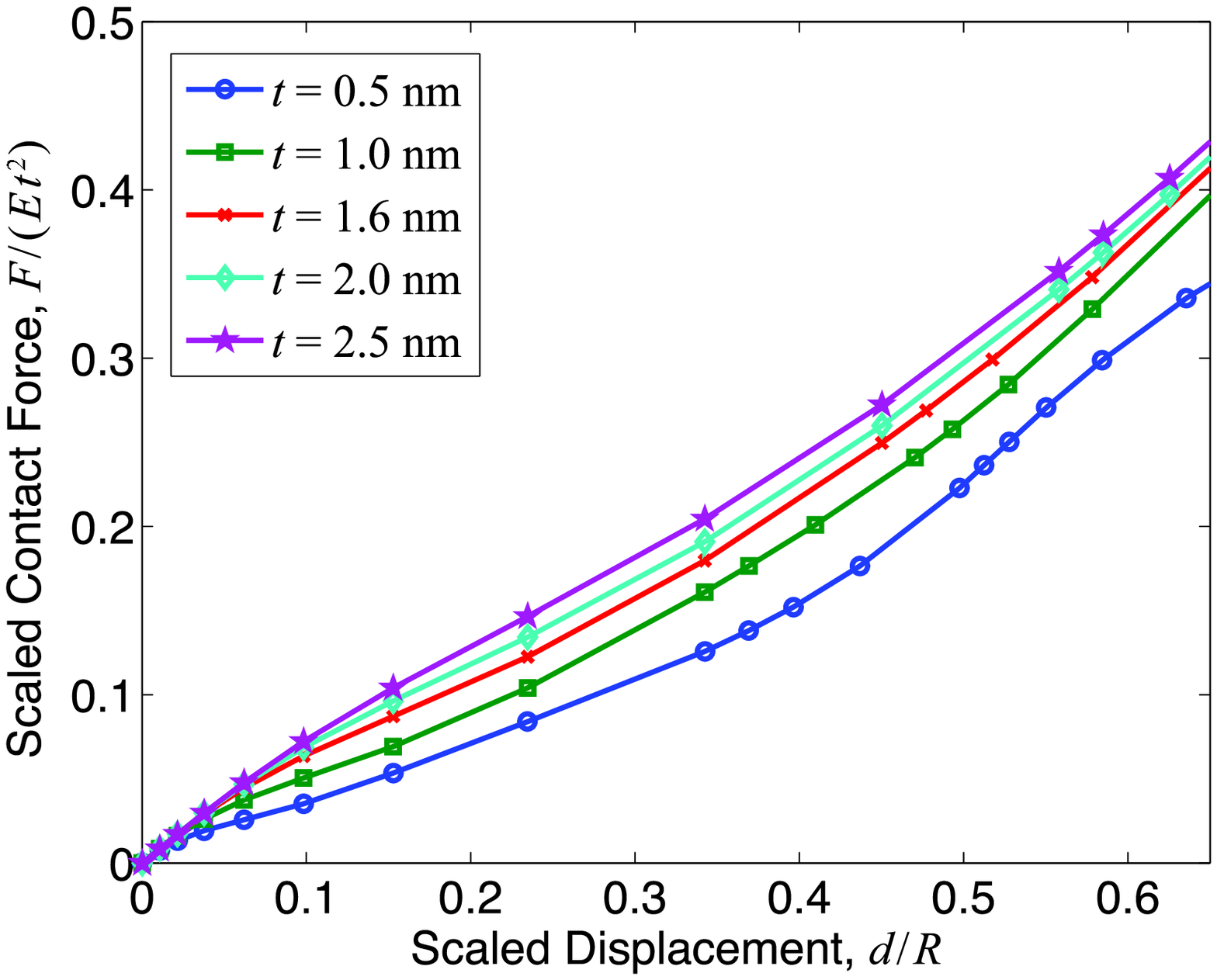}
  \caption{\label{fig:ForceComparisonPhi29}Contact force response of
    $\phi 29$ over a range of thicknesses, showing the dimensionless
    renormalization of the force according to expected scaling from
    linear elasticity.  The average of the two principal radii
    ($R=23.2$ nm) is used to scale the plate displacement.}
\end{figure}
\fi

As seen in the sequence of deformed shapes in
Fig.~\ref{fig:StressContour}, the progression of the von Mises stress
is localized to the upper and lower regions of the capsid which
undergo the highest deformation.  However, the upper region of the
capsid develops stress much more quickly, and with a higher magnitude.
Due to the lack of axisymmetry of the capsid about the indentation
axis, there is a region away from the region of contact with the AFM
tip where the stress begins to be prominent at $d \approx 17$ nm: the
inner surface of the upper portion of the capsid nearest to the
center.  However, this only occurs at displacements in the simulation
that exceed the displacements at which failure occured experimentally.
At the point when drops in the force are seen experimentally, $d
\approx 12$ nm, the highest stresses in the simulated capsid are
$\approx 760$ MPa for $E=4.5$ GPa.  This is on par with the ultimate
strength of typical protiens \citep{Howard:2001}, suggesting that
breakage of the capsid could very well be the cause of failure in
experiments.  The highest strain in the capsid is $\approx 11 \%$ at
$d\approx 12$ nm, and occurs on the inner surface of the capsid
directly under the AFM tip.  However, unlike the CCMV simulations, no
separation of the capsid away from the AFM tip is observed.

\ifFigsAtEnd \relax \else
\begin{figure}
  \centering
  \includegraphics[width=4in]{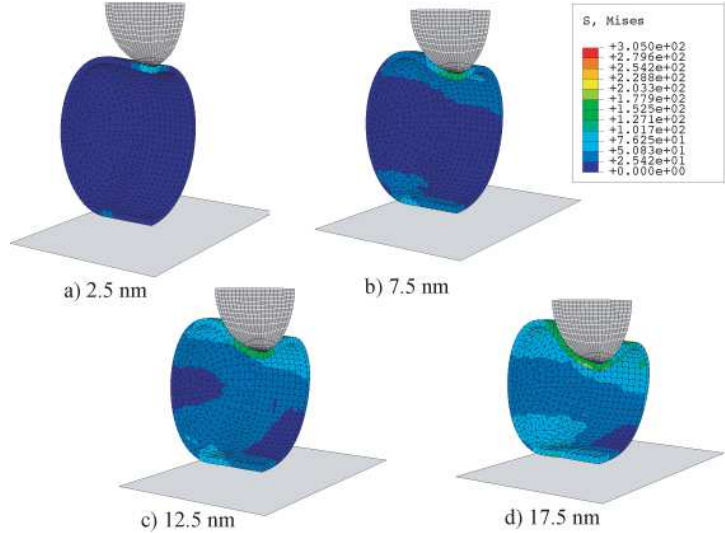}   
  \caption{\label{fig:StressContour} Indentation of ellipsocylindrical
    model of $\phi 29$ with dimensions from Fig. \protect\ref{fig:Schematic},
    $E=1.5$ GPa, and $\nu=0.4$.  Deformed capsid shapes are shown for
    several values of substrate displacement.  Color contours indicate
    the von Mises stress.}
\end{figure}
\fi

\section*{Discussion}\label{sec:discussion}

A core objective of this work has been to illuminate some of the
mechanisms responsible for the linear force-indentation responses
measured for the capsids of viruses CCMV and $\phi29$.  The
acknowledgment that these nanoscale structures act as elastic
shells prompts the question of how similar they are mechanically to
macroscopic shells, which are modeled well by continuum elasticity
theory.  It is not clear how to answer this question directly given
that a consise quantitative theoretical description of the local
mechanics of the individual constituent capsid proteins and their
interactions is currently out of reach.  In the absence of such
detailed constitutive understanding, experminents such as those
performed with atomic force microscopy \citep{IvanovskaEtAl:2004,
Michel} provide a coarse-grained probe to measure the global
structural response of capsid shells.  The continuum modeling in this
work serves as a convenient coarse-grained theoretical complement,
partially clarifying what features of the observed experimental
results are generic for shells, and what features require detailed
constitutive knowledge for explanation.

The results here demonstrate that the generally nonlinear shape of the
force-indentation response for a homogeneous spherical shell is
affected significantly by the dimensions of the indentor and the shell
itself.  In particular, stiffening at low force/indentation, similar
to Hertz response for solid bodies in contact, is characteristic of
the indentation response for thicker shells with dimensions like the
CCMV capsid.  Small AFM tips and small capsid thicknesses both
facilitate bending deformation, which generally softens structure at
higher indentations.  Large tip sizes and shell thicknesses can have
the opposite effect, leading to further stiffening of the force
response at larger indentations.  

Consideration of nonlinearites in modeling capsid nanoindentation has
an impact on the interpretation of experiements.  Previous linearized
elasticity models constructed by \citet{IvanovskaEtAl:2004} estimated
a Young's modulus for $\phi29$ of $1.8$ GPa.  This result appears
mainly to be a function of two things: a proportionality factor found
analytically for a spherical shell with point loads applied at the top
and bottom, and a finite element model also designed with point loads
with the capsid modeled as a geodesic ellipsoid.  Together, these
factors lead to an estimate of the Young's modulus which is smaller
than that of the present model.  Clearly such difference is to be
expected with the point loading exchanged for a finite-sized tip.
However, both the previously reported value and the value reported
here show the $\phi 29$ capsid to have a stiffness in the range of
$1-5$ GPa.

The continuum shell models presented here may also help to explain how the
linearity of the experimentally measured force-indentation responses
of $\phi29$ \citep{IvanovskaEtAl:2004} and CCMV \citep{Michel} can
persist despite the pervasiveness of nonlinearity in the system.  The
dimensions of the AFM tips and capsids in recent experiements on CCMV
are in a range in which the nonlinear stiffening and softening effects
seem to balance each other so that the modeled elastic
force-indentation response is close to linear.  This result suggests
that similar experiments done with AFM tips or capsids of other
dimensions might reveal force-indentation responses which are
\emph{nonlinear} even in the elastic regime.  However, though the
measured response of $\phi29$, which has significantly different
dimensions from CCMV, is also linear, the continuum models here still
show noticeable nonlinearity.  This indicates a deficiency of the
present continuum models for precise prediction of capsid
behavior. This could perhaps be remedied by a more accurate
representation of the geometry of the $\phi29$.  Alternatively,
the models could be improved with a more informed choice of
constitutive law, although the constitutive models employed here,
which are capable of significant nonlinearities, did not significantly
affect capsid shell response.

Another drawback to the isotropic, homogeneous, geometrically simple
model is that it does not produce drops in contact force that are
experimentally observed at $\approx 20-30\%$ deformation.  These drops
are signs of failure of capsid structure, which in general could be
triggered by geometric instability (buckling) or local failure of the
material (e.g., bond-breaking).  Over the relevant range of
dimensions, the models in this work did not exhibit any geometric
instabilities.  However, the present model does not take into account
any of the geometric complexities present in the molecular structure;
although spherical in an average sense, the CCMV capsid is faceted
into distinct regions of pentamers and hexamers and there are specific
and finite points where the subunits are joined through chemical
bonding.  The same is true for $\phi 29$.  These geometric
inhomogeneities could change the buckling characteristics of the
structure.  Also, the present model neglects the effects of
``pre-stressing'' consistent with the hypothesis of \citet{lidmar}
that the icosahedral vertices of spherical capsids act as
disclinations in an otherwise hexagonal protein lattice.  In fact,
based on this hypothesis, the recent coarse-grained molecular-dynamics
simulations of \citet{VliegenthartEtAll:2006} reveal that buckling, marked by
force drops, is indeed possible during capsid indentation.

Because the nonlinear elasticity model did not turn out to predict any
drop in force due to geometric instability, it may be inferred that the
mechanism of failure during the experiment is likely to be, at least
in part, capsid breakage.  To test this hypothesis, a more complex
model would be needed that describes the physics of interactions
between the subunits and the more variated structure of the capsid.  A
plausible scenario is that the capsid structure fails at the
noncovalent bonds between subunits.  It is possible that a true atomic
model would be required to capture points of failure
nucleation/initiation.  

\bibliography{capsids}

\ifFigsAtEnd 

\clearpage
\section*{Figure Legends}
\subsubsection*{Figure~\ref{fig:Schematic}.}
Schematic of idealized capsid models.  Dimensions are averaged from
the results of structural studies of CCMV \citep{ccmv} and $\phi29$
\citep{TaoOlson1998}.

\subsubsection*{Figure~\ref{fig:undeformedMesh}.}
Finite element meshes of CCMV (left) and $\phi 29$ (right) Capsids.
Dimensions of the capsids are shown in Fig.~\ref{fig:Schematic}.
AFM tip is modeled as a rigid hemispherical shell with $R=14$ nm,
substrate as a rigid flat plate. (Capsids are shown not on the same
scale.)

\subsubsection*{Figure~\ref{fig:StressContourCCMV}.}
Indentation of spherical model of CCMV with dimensions from Fig.~\ref{fig:Schematic}, and a constitutive response defined by the St.\
Venant-Kirchhoff theory with $E=250$ MPa, and $\nu=0.4$.  Deformed
capsid shapes are shown for several values of substrate displacement.
Color contours indicate the von Mises stress.

\subsubsection*{Figure~\ref{fig:CCMVforceResponse}.} 
Force response for indentation of spherical CCMV model from
Fig.~\ref{fig:StressContourCCMV}.

\subsubsection*{Figure~\ref{fig:TipSeparation}.}
At higher deformations, the apex of the CCMV capsid is observed to
separate from the indenting AFM tip.  Separation is preceded by a drop
in contact pressure near the apex.

\subsubsection*{Figure~\ref{fig:NeoHookean}.} 
Sensitivity to constitutive law.  Contact force curves for St.\
Venant-Kirchhoff, compressible Neo-Hookean, and compressible
Mooney-Rivlin models.  Representative capsid dimensions from
Fig.~\ref{fig:Schematic} are used.

\subsubsection*{Figure~\ref{fig:AFMTip}.}
Contact force comparison for several AFM tip sizes.  Representative
capsid dimensions from Fig.~\ref{fig:Schematic} are used.

\subsubsection*{Figure~\ref{fig:ForceComparison}.} 
Contact force response of CCMV over a range of capsid thicknesses.
(Top) Force vs. AFM plate displacement for $E=250$ MPa and
$R_{avg}=11.8$ nm. (Bottom) Dimensionless renormalization of force
response according to expected scaling from linearized thin-shell
elasticity.

\subsubsection*{Figure~\ref{fig:ForceComparisonPhi29}.}
Contact force response of $\phi 29$ over a range of thicknesses,
showing the dimensionless renormalization of the force according to
expected scaling from linear elasticity.  The average of the two
principal radii ($R=23.2$ nm) is used to scale the plate
displacement.

\subsubsection*{Figure~\ref{fig:StressContour}.} 
Indentation of ellipsocylindrical model of $\phi 29$ with dimensions
from Fig. \ref{fig:Schematic}, $E=1.5$ GPa, and $\nu=0.4$.  Deformed
capsid shapes are shown for several values of substrate displacement.
Color contours indicate the von Mises stress.

\clearpage

\begin{figure}
\centering
\includegraphics*[width=4in]{figure1.eps}
\caption{\label{fig:Schematic}}
\end{figure}

\clearpage

\begin{figure}
  \centering
  \includegraphics[width=4in]{figure2.eps}
  \caption{\label{fig:undeformedMesh}}
\end{figure}

\clearpage

\begin{figure}
  \centering
  \includegraphics[width=4in]{figure3.eps}
  \caption{\label{fig:StressContourCCMV}}
\end{figure}

\clearpage

\begin{figure}
\centering
\includegraphics[width=4in]{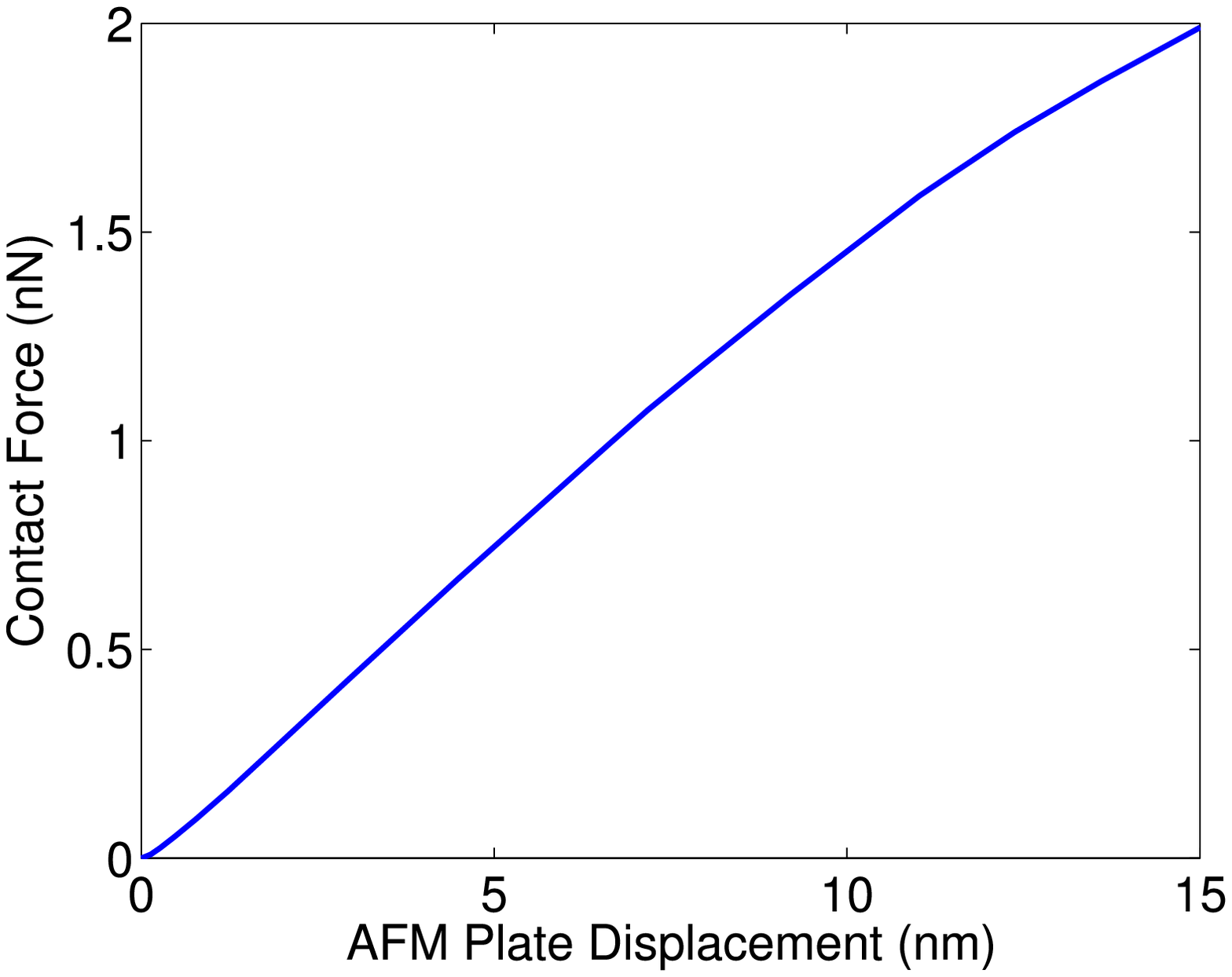}
\caption{\label{fig:CCMVforceResponse}}
\end{figure}

\clearpage

\begin{figure}
  \centering
  \includegraphics[width=4in]{figure5.eps}
  \caption{\label{fig:TipSeparation}}
\end{figure}

\clearpage

\begin{figure}
  \centering
  \includegraphics[width=4in]{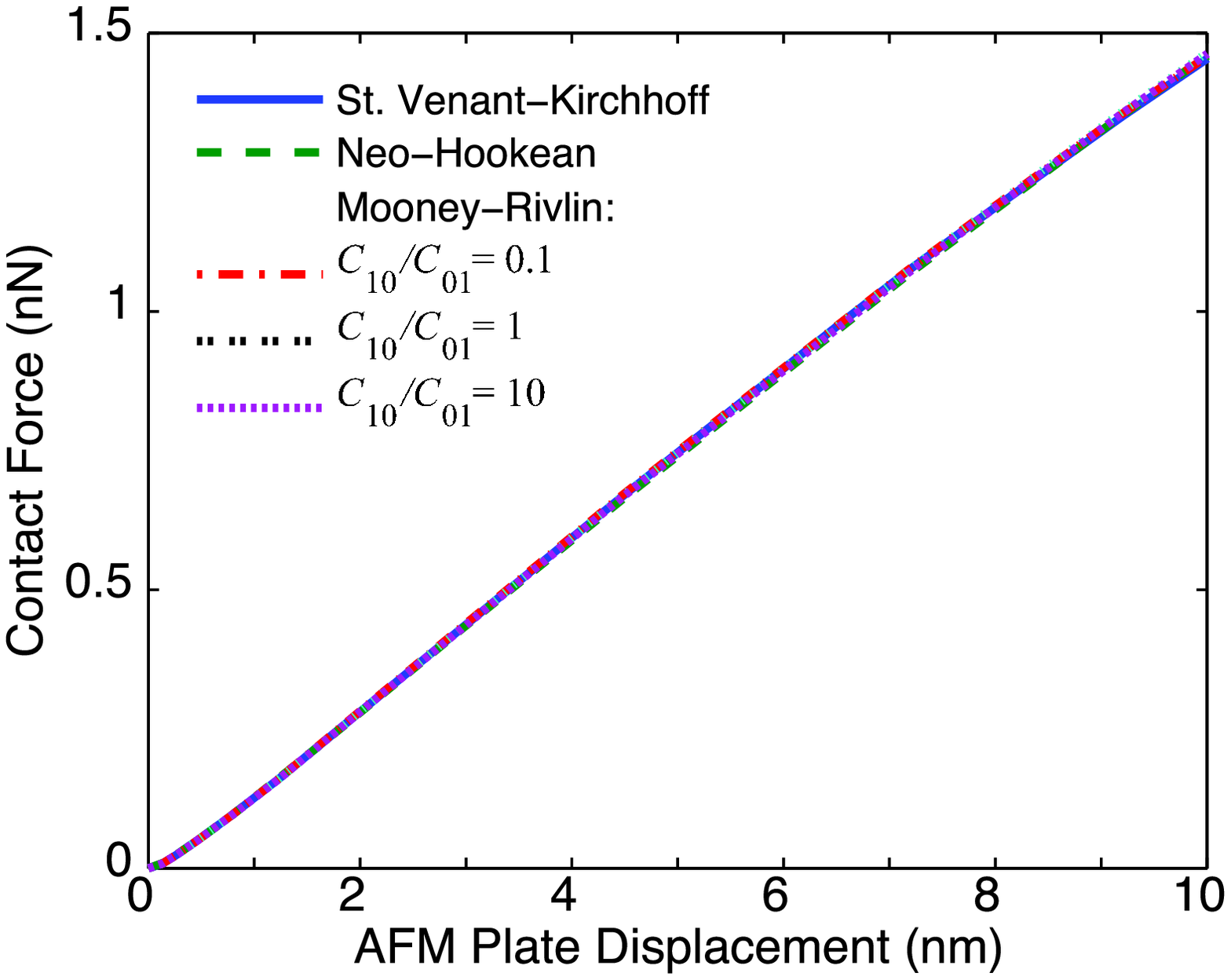}
  \caption{\label{fig:NeoHookean}}
\end{figure}

\clearpage

\begin{figure}
  \centering
  \includegraphics[width=4in]{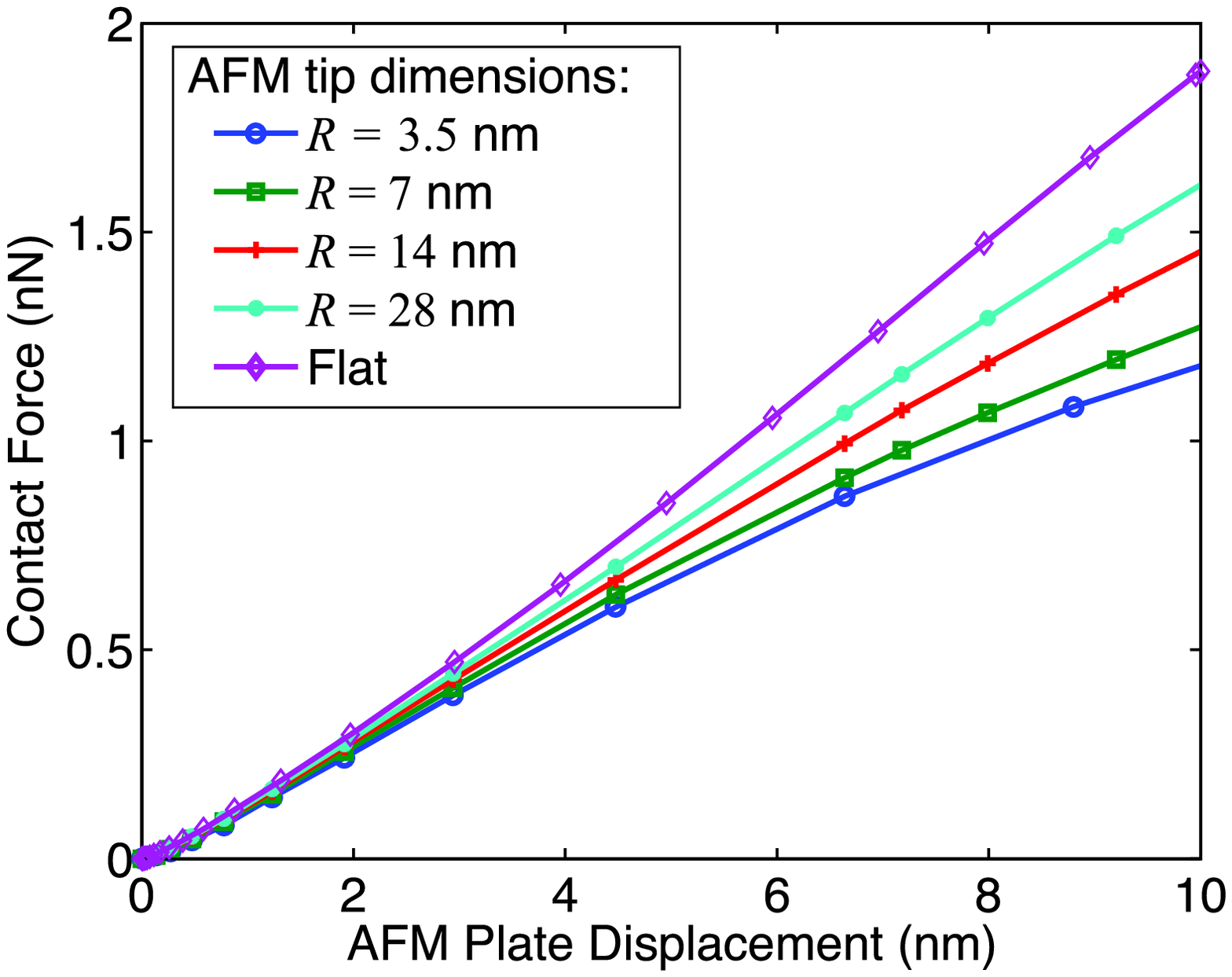}
  \caption{\label{fig:AFMTip}}
\end{figure}

\clearpage

\begin{figure}
  \centering
  \includegraphics[width=4in]{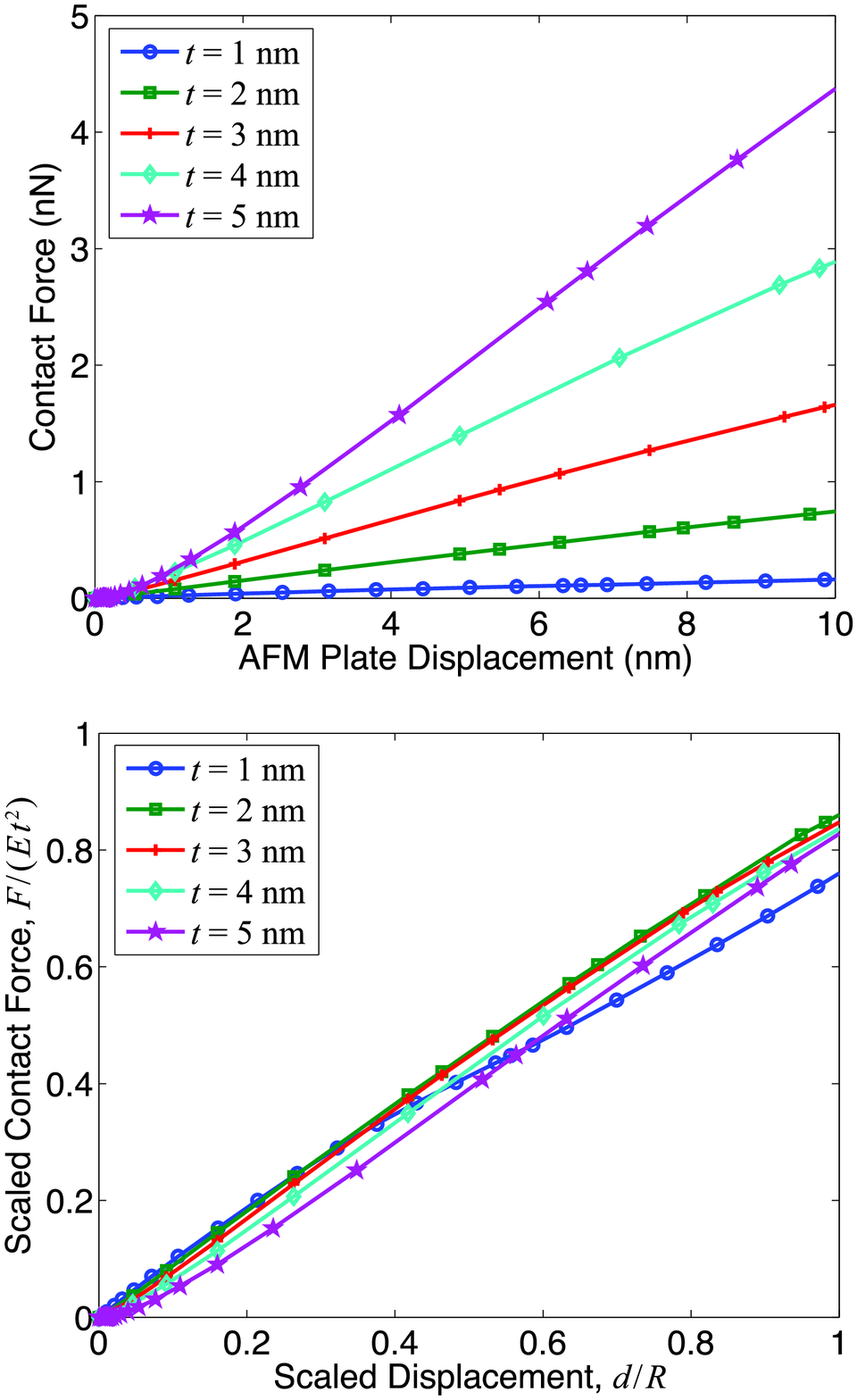}
  \caption{\label{fig:ForceComparison}}
\end{figure}

\clearpage

\begin{figure}
  \centering
  \includegraphics[width=4in]{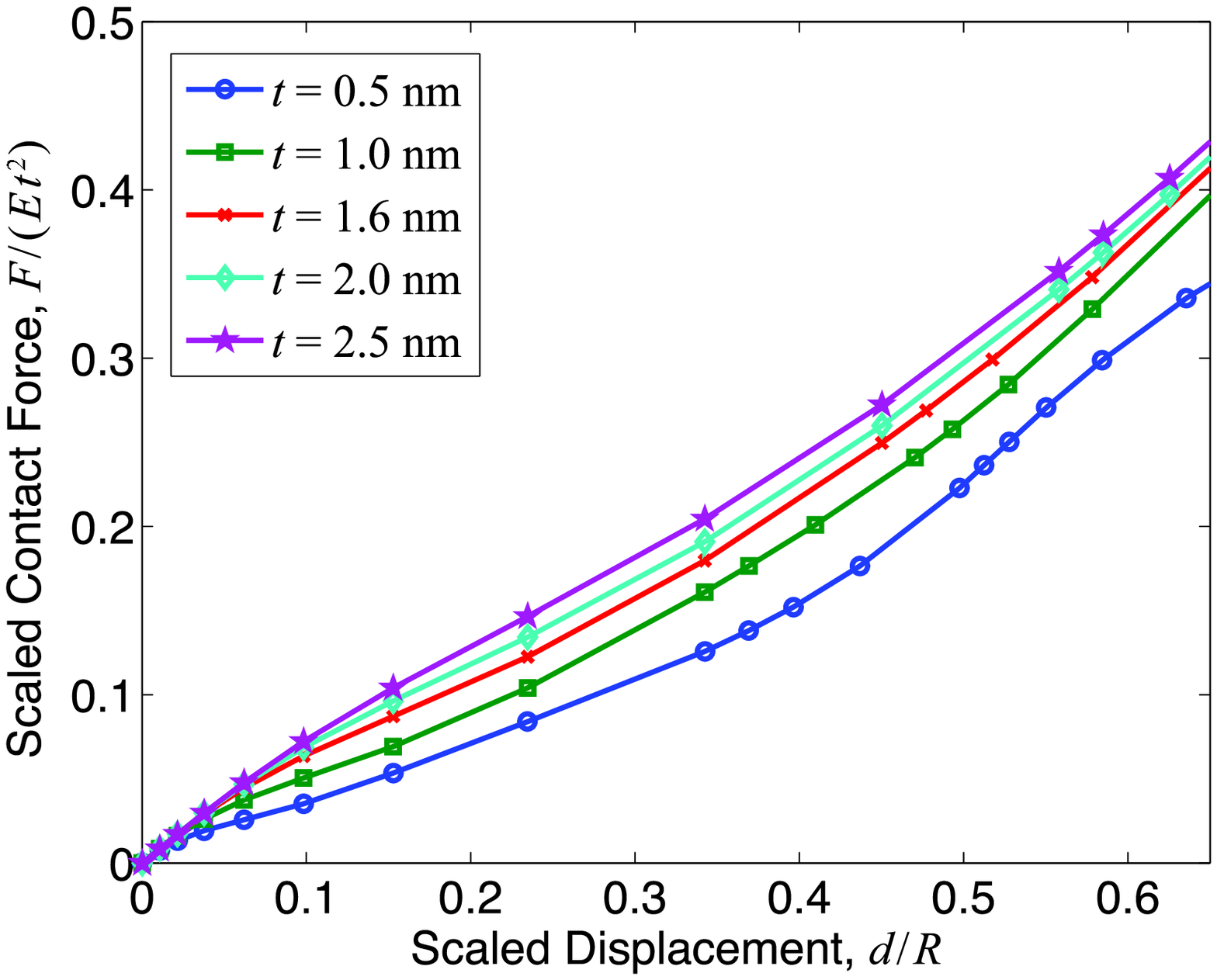}
  \caption{\label{fig:ForceComparisonPhi29}}
\end{figure}

\clearpage

\begin{figure}
  \centering
  \includegraphics[width=4in]{figure10.eps}   
  \caption{\label{fig:StressContour}}
\end{figure}

\fi

\end{document}